\documentclass[aps,prl,showpacs,twocolumn,longbibliography]{revtex4-1}
\usepackage{xcolor}
\usepackage{graphicx,dcolumn,bm}
\usepackage{amsfonts,amsmath,amssymb}
\usepackage{mathptmx}
\usepackage{longtable}
\usepackage{float}

\newcommand{\GG}{{\bf G}}
\newcommand{\Gp}{{\bf G}'}
\newcommand{\kk}{{\bf k}}
\newcommand{\kp}{{\bf k}'} 
\newcommand{\qq}{{\bf q}}
\newcommand{\rr}{{\bf r}}
\newcommand{\rp}{{\bf r}'}
\begin{document} 
\title{Superconductivity in CuCl/Si: possible excitonic pairing?}
\author{S. H. Rhim$^{1,2}$}
\email{sonny@ulsan.ac.kr}
\author{R. Saniz$^{3}$}
\author{M. Weinert$^{4}$}
\author{ A. J. Freeman$^{1,\dagger}$}
\affiliation{
$^1$Department of Physics and Astronomy, 
Northwestern University, Evanston, IL 60208, USA \\
$^2$Department of Physics and Energy Harvest Storage Research Center,
University of Ulsan, Ulsan, Republic of Korea\\
$^3$Department of Physics, University of Antwerp, Belgium\\
$^4$Department of Physics and Laboratory for Surface Studies,
University of Wisconsin-Milwaukee, Milwaukee, WI,53202, USA\\
$\dagger$ deceased
}
\date{\today}
\begin{abstract}
The search for superconductivity with higher transition temperature ($T_C$) has long been
a challenge in research efforts ever since its first discovery in 1911.
The effort has led to the discovery of various kinds of superconductors and progress in 
the understanding of this intriguing phenomenon.
The increase of $T_C$ has also evolved;
however, the dream of realizing room-temperature superconductivity is far from reality.
For superconductivity to emerge,
the effective quasiparticle interaction should overcome the repulsive
Coulomb interaction. This can be realized via lattice or spin degrees of freedom.
An alternative pairing mechanism, the excitonic mechanism,
was proposed 50 years ago,
hoping to achieve higher $T_C$ than by phonon mediation.
As none of physics principles has ever prevented excitonic pairing,
the excitonic pairing mechanism is revisited here
and we show that the effective quasiparticle interaction without lattice and spin
can be attractive solely electronically.
\end{abstract}
\maketitle  

For superconductivity to emerge,
the formation of Cooper pairs is necessary,
which requires an attractive quasiparticle interaction.
The Cooper pair is a bound state of two electrons. 
Two electrons on the Fermi surface with momenta $\kk$ and $-\kk$ scatter
to other states with momenta $\kp$ and $-\kp$ via
an attractive interaction, which is commonly called a binding glue.
The thermodynamic properties of a superconductor greatly depend on the kind of the glue,
for example phonons or spin-fluctuations.

According to BCS theory\cite{BCS},
the superconducting transition temperature ($T_C$) is expressed as
\begin{equation}
  \label{eq:tc}
  T_C \approx \Theta \cdot \exp\left(-1/\lambda\right),
\end{equation}
where $\lambda$ is the effective pairing strength,
and the prefactor $\Theta$ is the characteristic energy scale
of mediating bosons.
In phonon-mediated superconductors, 
$\Theta$ is the Debye temperature ($\Theta_D$),
and $\lambda$ arises from the electron-phonon coupling constant;
many attempts to raise $\lambda$ to obtain a higher $T_C$ often resulted in
a lattice instability (as in A15 compounds\cite{allen-mitrovic}).
Materials with high $\Theta_D$ usually have low $\lambda$.
At a time it was believed that $T_C \lesssim 30$K by phonon mediation\cite{Ginzburg70-1}.
Pairing mechanisms other than phonon with larger $\Theta$ has been been pursued in this context.

\begin{figure}[h]
  \centering
  \includegraphics[width=\columnwidth]{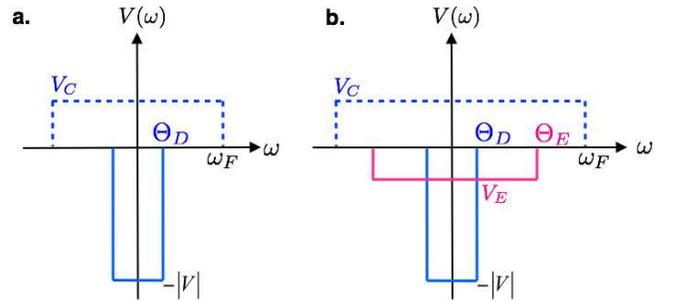}
\caption{(Color online) Schematic illustration of the effective potential in 
(a) BCS and Eliashberg theory, and (b) excitonic pairing ({\em or} ABB model).
$\Theta_D$ and $\omega_F$ are the Debye frequency and the Fermi energy, respectively,
$V_{C}=0$ in BCS theory.
$\Theta_E$ is the characteristic energy scale of excitons.
}
  \label{fig:1}
\end{figure}

Pairing by the spin-fluctuation mechanism relies
on spin-dependent interactions,
whose details are very sensitive to the crystal structure
and to the magnetic properties\cite{leggett75:rmp,monthoux07:nat}.
Despite the progress in several decades,
and hope that a spin-fluctuation mediation might offer a better mean than phonons
for achieving higher $T_C$,
an exact formulation of the spin-fluctuation has not been completed at the level
of BCS or Eliashberg theory\cite{eliashberg60:}.
Either phonon- or spin-fluctuation mediated, superconductivity requires 
an attractive effective interaction,
which, in principle, yields a gap equation that can be solved for $T_C$.
This, however, is a rather formidable task.
Instead, in the framework of the BCS approximation,
the quasiparticle interaction is treated as an attractive potential well
for frequencies less than $\Theta_D$ ($\Theta_D \ll \omega_F$),
while the Coulomb repulsion prevails 
over the entire frequency range, as illustrated in Fig.~\ref{fig:1}(a).
In the excitonic pairing, on the other hand, 
another attractive potential is added for frequency less than $\Theta_E$ ($\Theta_D \ll \Theta_E \ll E_F$),
as shown in Fig.~\ref{fig:1}(b)\cite{abb:73}.
Before we proceed, a brief history of the excitonic mechanism is first given.


{\em History of excitonic mechanism.}
In the early 1960's, Little\cite{Little64} and Ginzburg\cite{Ginzburg64} proposed
an alternative pairing mechanism, an excitonic mechanism,
where the mediating bosons were excitons {\em or} longitudinal plasmons,
which have a much higher characteristic energy scale than phonons,
so that $T_C$ can be greatly enhanced due to the larger prefactor in Eq.~(\ref{eq:tc}).
Specific geometries were proposed:
(i) a linear chain with a polarizer alongside [Little],
(ii) a layered system with an insulator-metal-insulator {\em sandwich} structure [Ginzburg].
Later, Allender, Bray, and Bardeen (ABB) studied
a metal-semiconductor interface more in detail\cite{abb:73}.
ABB showed that excitonic mediation can enhance $T_C$ without excluding the phonon mediation.
In ABB theory, Cooper pairs in the metallic region penetrate into the semiconducting gap region
and scatter back to the metallic region, during which virtual excitons are created and annihilated.
The potential well in ABB theory has an additional attractive part,
$V_{e}<0$ with cutoff frequency $\Theta_E$
much larger than $\Theta_D$, as is schematically illustrated in Fig.~\ref{fig:1}(b).
Inkson and Anderson (IA) objected to the ABB theory,
arguing that the effective interaction is of no help in pairing\cite{inkson:73}.
However, in reply to IA,
ABB questioned the validity of the dielectric function that IA used\cite{abb:reply}.
Subsequently, Cohen and Louie (CL) considered the ABB theory from IA's perspective\cite{Louie}.
CL also claimed that IA's dielectric function is too approximate and argued that 
an attractive interaction could exist. 
Using a rather simplified geometry, however, these authors could not conclude that there is pairing.
Both ABB and CL insisted that the inclusion of local-field effects
would clarify the debate between ABB and IA.
Zakharov {\em et al.} extendend CL's work to a Si-jellium-Si sandwich structure
with rigorous {\em ab initio} calculations of the dielectric function 
and showed that an attractive interaction could exist for certain frequencies and wave vectors
but there is no overall pairing\cite{zakharov}.
However, the introduction of jellium is rather artificial, which can be different
from the metallic state arising at the interface.

On the experimental side, following ABB, several efforts were made,
especially using the interface of Pb-based narrow gap semiconductors\cite{Miller:73,miller:76}.
This turned out to be fruitless despite some indications of superconductivity.
In the late 1970's, possible superconductivity in CuCl under hydrostatic pressure
had attracted the physics community,
which exhibited diamagnetism over 90 K\cite{cwchu}.
To explain possible superconductivity in CuCl, Abrikosov proposed ``metallic excitonium'',
an analogue to the electron-phonon mediation for $m_h/m_e \gg 1$:
The hole replaces the role of the nucleus, and
Cooper pairs are mediated via the exchange of the metallic excitonium giving a higher $T_C$
than with phonon mediation.
However, Abrikosov's idea was based on the incorrect belief at that time
that CuCl had an indirect band gap.
Despite numerous subsequent works,
superconductivity in CuCl remains speculative\cite{Vezzoli81,Collins_84}. 
Just before the discovery of superconductivity in cuprates\cite{Bednorz-Mueller},
another possible superconductivity was reported in the CuCl/Si [111] interface
with diamagnetism in the range of 60--150 K, an anisotropic magnetic response, and 
a resistivity drop by five orders of magnitude at 77 K\cite{Mattes85,Mattes89}.
However, the works on CuCl/Si was not followed up for long time and
the superconductivity of CuCl/Si still remains open\cite{sonny:2007}.

Rather recently,
superconductivity at an interface has been reported
in chalcogenides\cite{fogel01:prl,fogel02:prb} and oxides\cite{reyren07:sci},
although the pairing mechanism is not clear yet.
On the theoretical side, 
the possibility of the excitonic pairing has been raised in
metal-halide interfaces\cite{arita04:prb} and microcavities\cite{laussy10:prl}.
With advances in fabrication techniques, and increasing interest in interface phenomena,
revisiting superconductivity with excitonic pairing in CuCl/Si would be is of great interest.
\begin{figure}[htpb]
  \centering
  \includegraphics[width=1.\columnwidth]{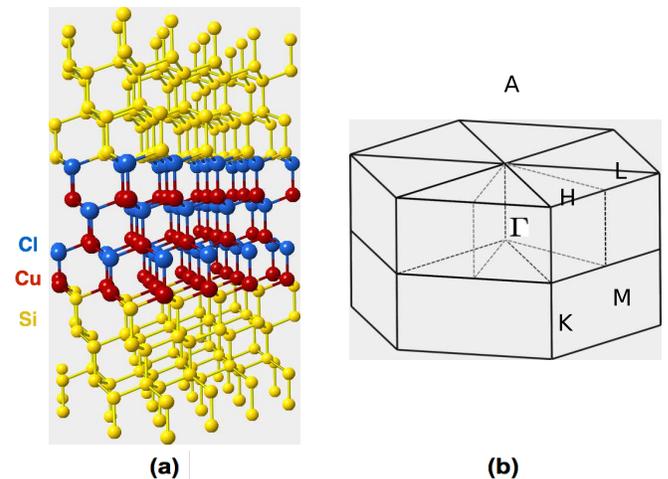}
  \caption{(Color online) (a) Structure of the CuCl/Si [111] superlattice for 
    $n=3$,where Cu, Cl, and Si atoms are denoted by red, blue, and yellow spheres,
    respectively.
   (b) Brillouin zone at high symmetry point directions.}
  \label{fig:2}
\end{figure}

{\em Electronic Structure.}
Density functional calculations are performed using the highly precise
full-potential linearized augmented plane wave (FLAPW) method\cite{wimmer:81}
with the local density approximation (LDA) for the exchange-correlation potential
by Hedin and Lundqvist parametrization\cite{hl:71}.
Muffin-tin (MT) radii of 2.15 a.u. (Cu) and 2.00 a.u (Si and Cl) are used,
where the angular momentum expansion $\ell \le 8$ is used inside the MT spheres.
Cutoffs for basis functions and potential representation
are 7.22 htr and 103.68 htr, respectively.
Structural optimization is done with a force criteria of 0.05 eV/$\AA$,
where the {\em in-plane} lattice constants are taken from experiments.
For summations in the Brillouin zone, 
a $12\times12\times4$ mesh is used for self-consistent calculations,
and $48\times48\times24$ mesh for Fermi surface plots. 
Results do not change when a denser {\em k} points mesh is used.


CuCl and Si crystallize in the zincblende and the diamond structure with lattice constants,
5.41~\AA~and 5.43~\AA, respectively, giving a very small lattice mismatch in the CuCl/Si superlattice.
The [111] direction of both the zincblende and the diamond structure are hexagonal with
an $ABC$ stacking sequence.
To ensure a periodicity along the $c$ direction,
(CuCl)$_n$/Si$_{4n}$ is taken into account, 
where $n$ denotes the number of CuCl layer.
The structure of the (CuCl)/Si superlattice for $n=3$ and
the corresponding Brillouin zone (BZ) with symmetry points are shown in Fig.~\ref{fig:2}.
Remarkably, the superlattice exhibits metallicity at interfaces
of the two semiconductors with appreciable band gaps of 3.4 eV (CuCl) and 1.14 eV (Si)\cite{sonny:2007}.
This metallicity is not due to the well-known LDA band gap underestimation,
since LDA gives finite band gaps for both CuCl and Si.
\begin{figure}[htpb]
  \centering
  \includegraphics[width=1.\columnwidth]{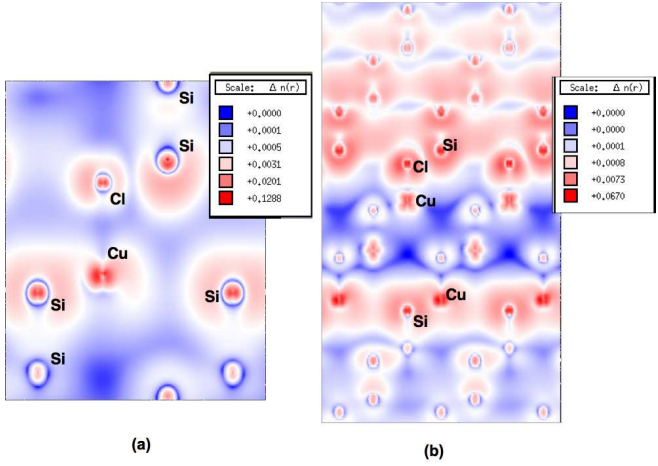}
  \caption{(Color online) Plot of charge densities at $\pm$0.25 eV
    with respect to $E_F$ (a) for $n=1$ and (b) $n=3$.
    Values in iso-plots are in logarithmic scale.
}
 \label{fig:3} 
\end{figure}

\begin{figure}[b]
  \centering
  \includegraphics[width=\hsize]{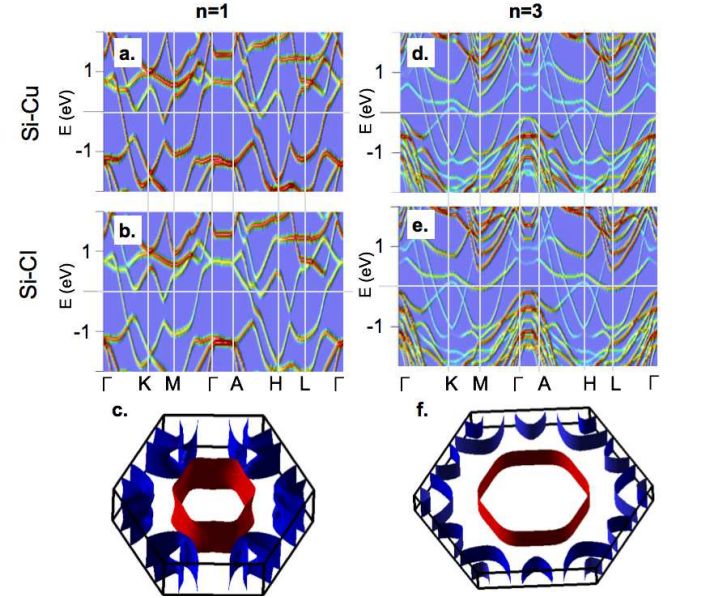}
  \caption{(Color online) Band structure of the CuCl/Si superlattice
    for $n=1$[(a)-(c)] and $n=3$[(d)-(f)].     
    Contributions from (a),(d) $n$-type and (b),(e) $p$-type interface.
    The Fermi surfaces for (c)$n=1$ and (f) $n=3$,
    where the electron- and hole Fermi surfaces are shown in red and blue, respectively.}
  \label{fig:4} 
\end{figure}

The metallicity at the interface is revealed in the charge density plot as shown in Fig.~\ref{fig:3},
where an energy slice of $\pm0.25$ eV with respect to $E_F$ is taken,
which correspond to density of 7.3$\times$10$^{21}$cm$^{-3}$ and 4.3$\times$10$^{21}$cm$^{-3}$ 
for $n=1$ and $n=3$, respectively.
The electronic structure also shows the interfacial metallicity [Fig.~\ref{fig:4}(a)-(d)],
where the $n$-type ($p$-type) metallicity is from the Si-Cu (Si-Cl) interface.
Furthermore, the Fermi surface (FS) evidences two-dimensionality with little dispersion along the z axis
at the zone center for the hole-FS and at the zone corner for the electron-FS.
As manifested in the plots of charge densities, bands, and Fermi surfaces,
the CuCl/Si superlattices exhibit interfacial metallicity,
which can be viewed as the insulator-metal-insulator sandwich structure, as Ginzburg proposed.
Hence, the possible superconductivity of this ``sandwich'' structure is here presented.

{\em Excitonic Superconductivity.}
In the presence of screening, the screened Coulomb interaction is expressed as\cite{hybertsen_Louie}
  \begin{eqnarray}
  \label{eq:Vscr}
  V^{s}_{\GG,\Gp}\left(\qq,\omega\right) &=& \frac{4\pi e^2}{|\qq+\GG|^2} \epsilon^{-1}_{\GG,\Gp}\left(\qq,\omega\right) \\
  V^{s}(\rr,\rp;\omega) &=& \sum_{\qq}\sum_{\GG,\Gp}
    e^{i\left(\qq+\GG\right)\cdot\rr} 
    \frac{4\pi e^2}{|\qq+\GG|^2}\\\nonumber
    &&\times\epsilon^{-1}_{\GG,\Gp}\left(\qq,\omega\right)
    e^{-i\left(\qq+\Gp\right)\cdot\rp}
\end{eqnarray}
where the inverse dielectric function is, 
within the random-phase approximation (RPA)
\begin{equation}
  \label{eq:invepsilon}
  \epsilon^{-1}_{\GG,\Gp}\left(\qq,\omega\right) = \delta_{\GG,\Gp} - \frac{4\pi e^2}{|\qq+\GG|^2}\chi_{\GG,\Gp}\left(\qq,\omega\right),
\end{equation}
and the generalized susceptibility, $\chi_{\GG,\Gp}\left(\qq,\omega\right)$ is given by
\begin{eqnarray}
  \label{eq:susceptibility}
&&  \chi_{\GG,\Gp}\left(\qq,\omega\right) = \frac{1}{\Omega}\sum_{\kk}\sum_{n,m}
\frac{ f_{\kk+\qq,n} - f_{\kk,m}}{ E_{\kk+\qq,n} - E_{\kk,m} - \omega +i\eta } \\ \nonumber
&\times&\langle \kk+\qq, n | e^{i\left(\qq+\GG\right)\cdot\rr}|\kk,m\rangle 
\langle \kk, m | e^{-i\left(\qq+\Gp\right)\cdot\rr}|\kk+\qq,n\rangle,
\end{eqnarray}
where $\Omega$ is the volume of the cell, {\em n,m} are band indices,
$f_{\kk,n}$ is the Fermi-Dirac function of the $n$-th band at $\kk$,
and $\GG,\Gp$ are reciprocal lattice vectors;
the {\em k} summation in Eq.~(\ref{eq:susceptibility}) is done over the full BZ.

We emphasize here that it is the inverse dielectric function, $\epsilon^{-1}\left(\qq,\omega\right)$,
rather than the dielectric function, $\epsilon\left(\qq,\omega\right)$,
that determines the screening in Eq.~(\ref{eq:Vscr}).
Moreover, it is the inverse dielectric function, $\epsilon^{-1}\left(\qq,\omega\right)$,
that is the true response function for any external perturbation
which satisfies the causality relation\cite{Kirznits,ginzburg77:book,Dolgov:rmp},
which leads to $0 < \epsilon^{-1}\left(\qq,\omega\right)<1$, or
$\epsilon\left(\qq,\omega\right)>1$ or $\epsilon\left(\qq,\omega\right)<0$:
The negative sign is allowed for $\epsilon\left(\qq,\omega\right)$
without violating the stability criterion\cite{Kirznits,ginzburg77:book,Dolgov:rmp}.

The pairing interaction
is formulated with time-reversal symmetry\cite{anderson59:jpcs}
and considering only singlet pairing,
in which a Cooper pair scatters from $(\kk,-\kk)$ to $(\kp,-\kp)$ in the presence of the screened interaction,
\begin{eqnarray}
  \label{eq:pairing_int}
  V^{p}_{\kk,\kp}\left(\omega\right) &=& \langle \kk,-\kk |~V^{s}(\rr,\rp;\omega)~|\kp,-\kp\rangle \\\nonumber
  &=& \sum_{\GG,\Gp}\sum_{n,m} 
  \frac{4\pi e^2}{|\qq+\GG|^2}\epsilon^{-1}_{\GG,\Gp}\left(\qq,\omega\right) \\\nonumber
  &\times&M_{n,m}\left(\kk,\qq,\GG\right)\cdot M^{\ast}_{n,m}\left(\kk,\qq,\Gp\right), 
\end{eqnarray}
where $M_{n,m}\left(\kk,\qq,\GG\right)=\langle \kk+\qq,n|e^{i\left(\qq+\GG\right)\cdot\rr}|\kk,m\rangle$
is the optical matrix.
[For a more detailed derivation, see Supplementary Information.]

%
\begin{figure}[t]
  \centering
  \includegraphics[width=0.9\columnwidth]{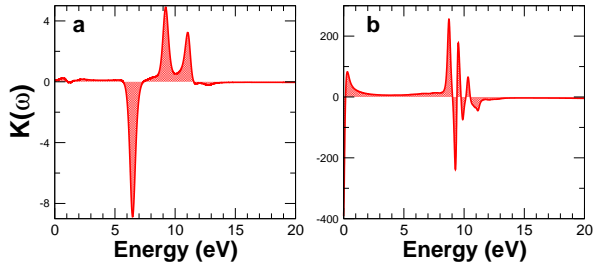}
  \caption{(Color online) Kernel function, $K\left(\omega\right)$, for (a) $n=1$ and (b) $n=3$
    CuCl/Si superlattice}
  \label{fig:5}
\end{figure}

It is important to note that having an attractive pairing, $V_{\kk,\kp}^{P} <0$,
for particular $k$ points and frequencies is not sufficient 
to ensure the pairing.
Instead, the Kernel function, $K\left(\omega\right)$,
the average of the pairing interaction over the BZ zone, should be attractive.
The concept of the Kernel function was first introduced by Cohen
when he proposed superconductivity in doped semiconductors\cite{Cohen:64}
and it is extended in Lithium under high pressure by Akashi and {\em et al.}\cite{akashi13:PRL}.
Here we rewrite the Kernel function as
\begin{equation}
  \label{eq:kernel}
  K\left(\omega\right) = \sum_{\qq}K_{\qq}\left(\omega\right),
\end{equation}
where $K_{\qq}\left(\omega\right)$ is $\qq$ decomposed Kernel function.
[For derivation of the Kernel function, see the Supplementary Information.]

The calculated $K\left(\omega\right)$ of the CuCl/Si superlattices are shown in Fig.~\ref{fig:5} for
$n=1$ and $n=3$, respectively.
Indeed, regions of $K\left(\omega\right)< 0 $ are found, a clear indication that pairing can exist
due to the electronic interactions alone without involving lattice or spin degrees of freedom.
The attractive interaction occurs around 6 eV and 9 eV for $n=1$ and $n=3$, respectively.
While the attractive region of $n=1$ is rather wide, that in $n=3$ is narrow.
In both cases, regions of highly repulsive interaction are present nearby the attractive region.
However, the presence of the repulsive region of $K\left(\omega\right)$ does not necessarily prevent pairing.
In phonon mediation,
pairing occurs in a very thin range of energy ($\Theta_D$) in the Fermi surface.
At least in $s$-wave pairing symmetry,
the presence of attractive region in $K\left(\omega\right)$ favors pairing.

\begin{figure}[bhtp]
  \centering
  \includegraphics[width=1.0\columnwidth]{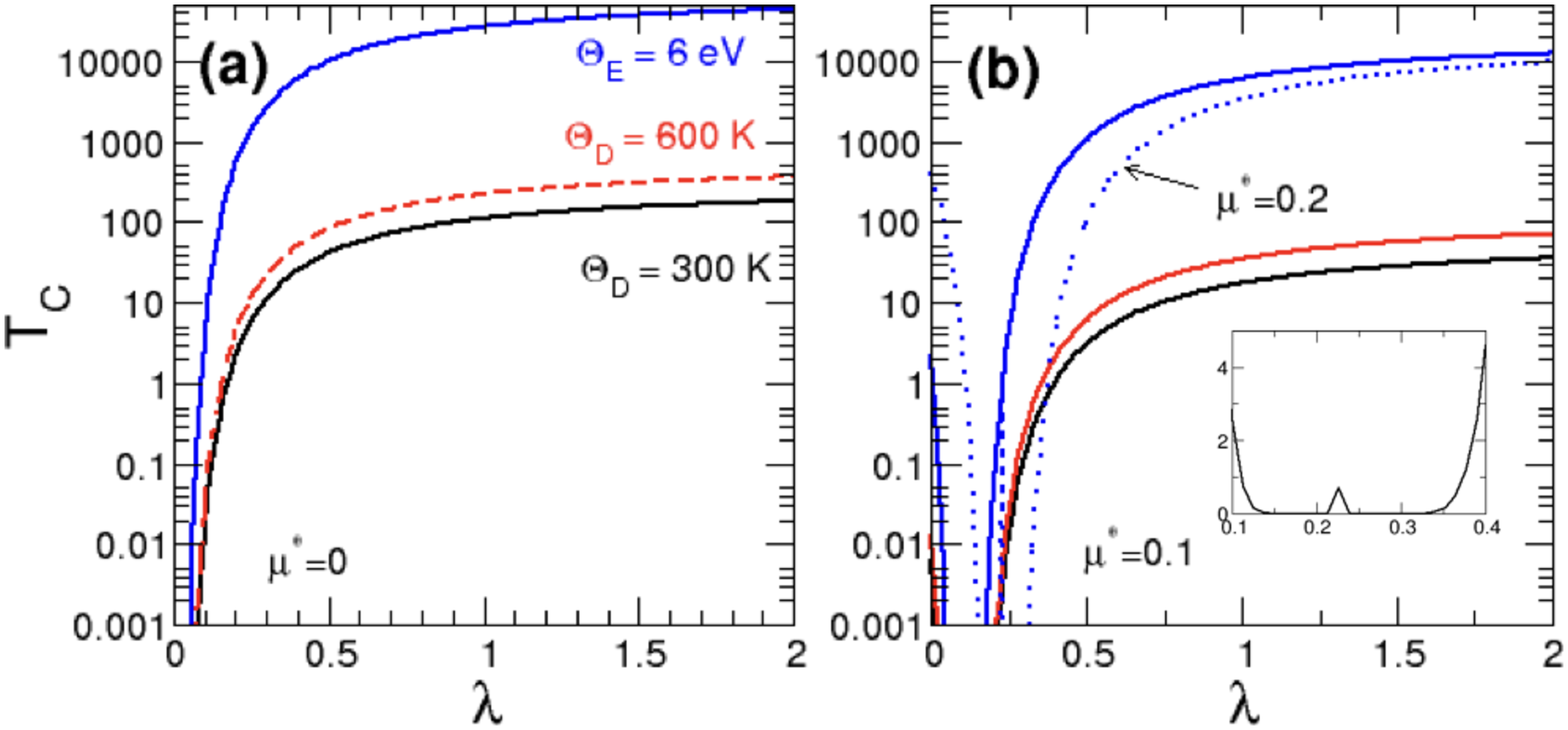}
  \caption{(Color online) $T_C$ (in logarithmic scale) as function of $\lambda$ for 
    (a) without Coulomb repulsion using Eq.~(\ref{eq:tc})
    and (b) with Coulomb repulsion, $\mu^{\ast}$ =0.1, using McMillan's equation\cite{McMillan}.
    $T_C$ with excitonic pairing with $\Theta_E=$ 6 eV is shown in blue,
    phonon pairing with $\Theta=$300 and 600 K are shown in black and red dotted lines,
    respectively.
    For simplicity, 
    the asymptotic behavior of $T_C$ for a large value of $\lambda$ is ignored.
    The electron-exciton pairing constant,
    $\lambda$, here is defined as done for the electron-phonon coupling constant.
    The case for $\mu^{\ast}$=0.2 is shown in dotted line in (b).
    Inset: $T_C$ (in linear scale) for $\mu^{\ast}$=0.2 near $\lambda$ between 0.1 and 0.4.
    }
  \label{fig:6}
\end{figure}
%
%
$T_C$ as function of $\lambda$ is plotted in Fig.~\ref{fig:6}.
$\Theta_E$ is approximated to 6 and 9 eV from Fig.~\ref{fig:5} for $n=1$ and $n=3$, respectively.
Fig.~\ref{fig:6}(a) is without Coulomb repulsion ($\mu^{\ast}$) using Eq.~(\ref{eq:tc}),
where phonon pairing with $\Theta_D$=300 and 600 K are shown for comparison.
One notes that $T_C$ by excitonic pairing with $\mu^{\ast}=0$ for $\lambda>0.5$ become unrealistically high.
The case with nonzero $\mu^{\ast}$ needs some caution. $T_C$ with $\mu^{\ast}=0.1$ is shown in Fig.~\ref{fig:6}(b),
where McMillan's equation is used. As well expected, $T_C$ is reduced with inclusion of $\mu^{\ast}$. 
Even when $\mu^{\ast}=0.2$ [dotted line], enhancement of $T_C$ is evident despite reduction compared to Fig.~\ref{fig:6}(a).

In this $T_C$ estimate, the validity of Migdal theorem, whether to include higher order vertex correction,
is ignored for simplicity. 
It has been argued that $T_C$ reduces by vertex correction\cite{rietschel83:prb}.
On the other hand, enhancement of $T_C$ in layered systems has been proposed
even with inclusion of vertex correction\cite{BMK03:prb,atwal04:prb}.
As Coulomb repulsion ($\mu$) is renormalized to 
$1/\mu^{\ast}=1/\mu + \log\left(\frac{\omega_F}{\Theta_D}\right)$ in phonon mediation\cite{Morel-Anderson},
in the excitonic pairing, $\Theta_E$ replaces $\Theta_D$ hence $\mu^{\ast}$ will be larger than phonon mediation.
Nevertheless, as shown in Fig.~\ref{fig:6}, $T_C$ enhancement by exciton mediation is evident\cite{comment}.

Recall that the diamagnetism of CuCl/Si [111] was reported to be in the 60--150 K range,
and calculations within the electron-phonon mediation gave $T \leq 2$K\cite{sonny:2007}for this system;
thus if CuCl/Si [111] is indeed superconducting, it cannot be solely by phonon mediation.
We want to point out here that exciton pairing is not the sole pairing mechanism of purely electronic in character.
Kohn and Luttinger\cite{kohn65:prl} suggested that purely electronic pairing
due to the sharp edge of the Fermi surface can be realized via a Friedel-like density oscillation,
which enables an attractive interaction, whose $T_C$, however, is too low to be observed experimentally.

{\em Summary}
In summary, the CuCl/Si superlattice was revisited.
It clearly shows metallicity at interfaces, as evidenced
in bands, Fermi surfaces and charge density plots. 
The possibility of excitonic pairing is shown by calculating the Kernel function $K\left(\omega\right)$.
The attractive effective interaction can exist for frequencies ranging over several eV
in favor of pairing with large prefactor in the $T_C$ equation.
In spite of renormalization, which would increase $\mu^{\ast}$ in the exciton pairing,
enhancement of $T_C$ over the phonon mediation is evident.
%
%

  We are grateful to J. B. Ketterson and P.R. Sievert
  for their encouragement and fruitful discussions, and to
  the early encouragement of J. Bardeen.
  This work was supported by the U.S. Department of Energy (DE-FG02-05ER45372) and
  a time grant by DGIST supercomputing center.
  Work at Ulsan is supported by the Priority Research Centers
  Program (NRF-2009-0093818) and the Basic Science Research Program (NRF-2015R1A2A2A01003621)
  funded by the Ministry of Education and the Ministry of Science, ICT, and Future Planning.

%

\begin{thebibliography}{43}
\bibitem [{com()}]{comment}%
  \BibitemOpen
  \href@noop {} {}\bibinfo {note} {In Fig.6(b), singularity near $\lambda \sim
  \mu^{\ast}$ is associated with argument of exponential in McMillan's
  equation.}\BibitemShut {Stop}%
\end{thebibliography}
\end{document}